\documentclass[pra,twocolumn,nofootinbib,letter]{revtex4}
\usepackage[english]{babel}
\usepackage{graphicx}
\usepackage{bm,amsmath}
\usepackage{dcolumn,longtable}
\usepackage{natbib}
\usepackage[colorlinks]  
{hyperref}
\usepackage{float}
\usepackage{wrapfig}

\newcommand{\fref}[1]{Fig.~\ref{#1}}

\newcommand{\sref}[1]{Sec.~\ref{#1}}

\newcommand{\tref}[1]{Table~\ref{#1}}

\begin{document}

\title{Prediction of quantum many-body chaos in protactinium atom}

\author{A. V. Viatkina$^{1}$}
\author{M. G. Kozlov$^{2,3}$}
\author{V. V. Flambaum$^{1,4}$}
\affiliation{$^1$Helmholtz Institute, Johannes Gutenberg University, 55099 Mainz, Germany}
\affiliation{$^2$Petersburg Nuclear Physics Institute, Gatchina 188300, Russia}
\affiliation{$^3$St.\ Petersburg Electrotechnical University ``LETI'', Russia}
\affiliation{$^4$School of Physics, University of New South Wales, Sydney 2052, Australia}
\date{
\today}

\begin{abstract}
Energy level spectrum of protactinium atom (Pa, $Z=91$) is simulated with a CI calculation. Levels belonging to the separate manifolds of a given total angular momentum and parity $J^\pi$ exhibit distinct properties of many-body quantum chaos. Moreover, an extremely strong enhancement of small perturbations takes place. As an example, effective three-electron interaction is investigated and found to play a significant role in the system. Chaotic properties of the eigenstates allow one to develop a statistical theory and predict probabilities of different processes in chaotic systems.

\end{abstract}

\maketitle

\section{Introduction}\label{sec_intro}
Spectra of complex many-body quantum systems often demonstrate universal statistical behavior. In the 1950-s Wigner showed that it can be modeled by means of random matrix theory (RMT) applying this method to atomic nuclei \cite{Wig57conf}. Soon after Rosenzweig and Porter published an analysis of experimental spectra of atoms \cite{RP60}:
several sixth-period transition metals ($Z= 72,\dots,77$) displayed an agreement with RMT in the nearest neighbor spacing distribution of their even-parity energy levels, whereas spectra of lighter atoms with similar electronic structure (transition metals $Z= 21,\dots,28$ and $Z= 39,\dots,46$) showed more regular behavior. This fact was attributed to applicability of LS-coupling scheme in lighter atoms. In 1983 it was demonstrated \cite{CG83} that experimental spectra of neutral atoms and ions of Nd, Sm and Tb follow predictions of RMT.
Later a realistic numerical model of Ce was investigated \cite{FGGK94,FGGP98,FGGP99,GFG95} and it was shown that properties of its excited states are consistent with the behavior of random two-body interaction matrices \cite{FIC96, FGI96, FI97}.

Many-body systems that exhibit such properties are often called \textit{chaotic}. They are sensitive to small perturbations and for that reason extremely difficult to model accurately, since a small addition to the Hamiltonian results in a significant change of the energy levels. Chaotic properties of the eigenstates have important consequences. Chaos allows one to develop statistical theory and calculate matrix elements of different operators between extremely complex many-body states including electromagnetic transition probabilities and probabilities of other processes - see e.g. \cite{FG95, FG2000, FGG96, FGG98, FGGH02, DFGH13, FKG15}.

In fact, small perturbations in these systems are subject to \textit{statistical enhancement} due to the large number of principal basis components $N$ participating in an eigenfunction of a chaotic system \cite{SF80_en, SF82, Fla93, FV93}. Mixing of neighboring eigenstates $|\Psi_a\rangle$ and $|\Psi_b\rangle$ by a small single-particle interaction $V$ scales as:
\begin{equation}\label{sqrtN}
\frac{\langle\Psi_a|V|\Psi_b\rangle}{\Delta E_{ab}}\sim \sqrt{N},
\end{equation}
where $\Delta E_{ab}$ is the difference in energies between the states.

Eigenfunctions of compound nuclei tend to have $N\sim 10^4-10^6$ \cite{BM98_book}. Enhancement of parity-nonconserving effects for nuclei was predicted \cite{SF80_en, SF82} and subsequently measured \cite{ABV83} (see also review \cite{FG95} and references therein). The eigenfunctions of highly chaotic Ce atom were estimated to have $N\sim 10^2$. We show that protactinium has an order of magnitude higher $N\sim 10^3$. It leads to extremely strong enhancement of small perturbations, not unlike that in compound nuclei. Effective three-electron interaction, usually small in atoms, becomes remarkably strong in Pa, mixing the basis states and altering the positions of energy levels.


\subsection{Random matrices}\label{sec:rmt}
Consider the basic version of RMT of ensembles of matrices $N\times N,\ N\rightarrow\infty$ with Gaussian random elements, where each matrix follows a set of symmetry rules; the probability density of matrix to appear in an ensemble is determined by its trace. There are three most common ensembles: Gaussian Orthogonal (GOE), Unitary (GUE) and Symplectic (GSE) \cite{Dys62}. GOE is connected with Hamiltonians of time-reversal and rotationally invariant systems (or with systems without rotational invariance, but with integer spin); GUE is relevant in more general case when the time-reversal symmetry is broken; GSE is used for time-reversal invariant systems with half-integer spin and broken rotational symmetry \cite{GMW98}. 

The characteristic property of distribution of the eigenvalues in any of the named ensembles is the repulsion of neighboring levels. It is the strongest in GSE and the weakest in GOE.
By the presupposition of ergodicity, statistical properties of spectra of matrices across the ensemble are transferable to the spectrum of one of the wide range of matrices from the ensemble. 

The tool mostly used to examine the repulsion of levels is the nearest neighbor spacing (NNS) distribution. In matrices it is defined as follows: let $H$ be a matrix from one of the three ensembles, its eigenvalues listed as $E_1\leq E_2\leq \dots\leq E_N$. Take some of them $E_n\leq\dots\leq E_{k}$ and let them fall into a sufficiently large interval $\Delta E$. The spacings $S_i=E_{i+1} - E_{i}$ should be then divided by the average spacing $D$ within $\Delta E$ to receive the dimensionless $s_i$, which can be later compared with similarly normalized spacings from other parts of the spectrum.
\begin{align}
S_i &= E_{i+1} - E_{i}\\
D &= \langle S_i\rangle_{\Delta E}\\
s_i &= S_i/D
\end{align}
The probability for a normalized spacing $s_j$ to fall into an interval $[s, s+ds]$ is $dP=P(s)ds$ and the NNS distribution is then defined as the probability density $P(s)$. The procedure of obtaining dimensionless spacings $s_i$ from a spectrum non-uniform in density is called \textit{unfolding} \cite{Haake_book}. It can be performed either as shown above, through finding the average spacing on a limited-length interval and then moving the interval along the spectrum; or it is possible to derive local average spacing from a polynomial fit of the spectrum cumulative function. The latter method will be described below in \sref{nns_distr}.

Considering a two-dimensional case, Wigner predicted the NNS distribution of GOE to be of shape \cite{Wig57conf}:
\begin{equation}\label{eq:wigner_surmise_goe}
P_{\mathrm{GOE}}(s)=\frac{\pi s}{2}\mathrm{exp}\left(-\frac{\pi}{4}s^2\right),
\end{equation}
which was later named \textit{Wigner surmise}.
It turned out to be very close to the exact NNS distribution $p(s)$ for GOE calculated later \cite{Meh60,Meh04book}.

Along with the eigenvalues of matrices, the NNS distribution can be found for a large number ($N\rightarrow\infty$) of randomly and independently placed points on a limited interval \cite{BR06}. In this case the repulsion of neighboring points is absent; in fact, they tend to cluster. This $p(x)$ is referred to as \textit{Poisson NNS distribution}:
\begin{equation}\label{eq:poisson}
P(x) = e^{-x}.
\end{equation}

If the investigated system has good quantum numbers, its Hamiltonian matrix can be written in a block-diagonal form. The spectrum is then composed of non-interacting subsets of levels and its NNS may resemble Poisson \eqref{eq:poisson} more than the Wigner case \eqref{eq:wigner_surmise_goe} due to the absent repulsion.

It is useful to introduce one-parameter \textit{Brody function} \cite{Bro73}, which turns into Poisson distribution \eqref{eq:poisson} for $\eta=0$ and is close to Wigner distribution for GOE \eqref{eq:wigner_surmise_goe} when $\eta=1$:
\begin{align}\label{eq:brody}
P_\eta(s) &= As^\eta \mathrm{exp}(-\alpha s^{\eta + 1}),\\
A &= (\eta +1)\alpha,\\
\alpha &= \left[\mathrm{\Gamma}\left(\frac{\eta + 2}{\eta + 1}\right)\right]^{\eta+1}.
\end{align}
Thus we define the \textit{repulsion parameter} $\eta\in[0,1]$.

\section{Method}
\subsection{CI model}\label{ci_model}
Protactinium (Pa) is an actinide with atomic number $Z=91$. Its ground state has total angular momentum $J=11/2$, parity $\pi=+1$ and it belongs to the configuration [Rn]5f$^2$6d$^1$7s$^2$. The unfilled 5f shell along with five valence electrons gives rise to a complex and dense spectrum.

We use a CI package described in \cite{KPST15} to model overall statistical properties of lower energy levels of Pa. Hartree-Fock-Dirac one-electron functions $\phi_i$ are generated for the configuration [Rn] 5f$^2$6d$^1$7s$^2$7p$^0$. The $\phi_i$ of valence electrons are built in the field of frozen [Rn] core. They are arranged into Slater determinants $|\Phi_i\rangle$ belonging to 107 even or 100 odd relativistic configurations. For the basis of the Hamiltonian matrix we choose $|\Phi_i\rangle$ with the projection of total angular momentum $M=0.5$ to account for states with all possible $J$. 
We diagonalize the matrix $H$ and obtain the eigenfunctions and corresponding eigenvalues:
\begin{equation}\label{eigenfunct}
|\Psi_i\rangle = \sum_k C_{ik}|\Phi_k\rangle\ ,
\end{equation}
\begin{equation}
\hat{H}|\Psi_i\rangle = E_i|\Psi_i\rangle\ .
\end{equation}
The resulting energy spectrum is then split into subspaces of fixed total angular momentum and parity $J^\pi$, which are later analyzed separately.

Predictions of the model are compared with the experimental data \cite{BW92}. The straightforward CI calculation described above produces correct ground state and a plausible order of energy levels' leading configurations.

Strictly speaking, in the case of heavy open-shell atoms, we are not working with the pure RMT \cite{GFG95, FGGK94, FIC96, FGGP99}. Due to the two-body nature of residual Coulomb interaction, the matrix element $\langle\Phi_i|H|\Phi_j\rangle$ is zero when the basis determinants differ in more than two single-electron $\phi_i$. If the basis states $|\Phi_i\rangle$ are enumerated according to their energy $\varepsilon_k=\langle\Phi_k|H|\Phi_k\rangle$, then the matrix $H$ has diagonal consisting of ordered $\varepsilon_k$ and sparsely distributed off-diagonal elements $H_{ij}$, decreasing with larger distances $|i-j|$. Characteristic distance of this decrease is denoted as $b$, roughly corresponding to the \textit{bandwidth} of band random matrix theory (bRMT). Nevertheless, spacings of eigenvalues of such a matrix should follow the Wigner distribution \eqref{eq:wigner_surmise_goe} \cite{FIC96, FW71, BF71a}.

\subsection{Unfolding procedure}\label{nns_distr}
In order to bring the local density of the spectrum to unity, one needs to perform \textit{unfolding} \cite{Haake_book, GMW98}. Then for each subspace $J^{\pi}$ the unfolded NNS statistics is built and fitted with Brody function \eqref{eq:brody} to obtain the repulsion parameter $\eta$. 

We plot the cumulative function $N(E)$, where $N$ is the successive number of a level and $E$ its energy. The overall density of the spectrum defined as
\begin{equation}
\rho(E')=\sum_N \delta(E'-E_N)
\end{equation}
is connected to the cumulative function:
\begin{equation}
N(E)=\int_{-\infty}^{E}\rho(E')dE'.
\end{equation}
We approximate $N(E)$ with a fifth-order polynomial $p_{N(E)}(E)$ and find the smoothed form of the level density as its derivative:
\begin{equation}\label{eq:rho_smooth}
\rho(E)\equiv \rho_\mathrm{smooth}(E)=\frac{dp_{N(E)}}{dE}.
\end{equation}
This density can be understood as $\rho(E)=D^{-1}(E)$, where $D$ is the local mean level spacing. To build a NNS statistics we divide each spacing by the relevant $D(E)$:
\begin{equation}
s_i=\frac{S_i}{D}=S_i\rho.
\end{equation}
A set of unfolded dimensionless spacings $s_i$, $i=0, 1, 2,\dots, n$ is obtained.

\subsection{Strength function}
In order to establish the approximate number of basis states $|\Phi_k\rangle$ strongly participating in a given eigenfunction \eqref{eigenfunct} we investigate values of $C_{ik}$. Let the energy of a basis state (determinant) be defined as $\varepsilon_k=\langle\Phi_k|H|\Phi_k\rangle$. We enumerate the basis states according to their energy and plot the squared coefficients of an eigenfunction $|\Psi_i\rangle$ on the determinants energy axis as $|C_{ik}|^2=|C_{i}(\varepsilon_k)|^2$. Significantly large $|C_{i}(\varepsilon_k)|^2$ usually appear around the eigenvalue $E_i$ within a certain interval:
\begin{equation}
|E_i - \varepsilon_k|\lesssim\Gamma,
\end{equation}
where $\Gamma$ is called \textit{spreading width}. It is intimately connected to the Wigner \textit{strength function} \cite{Wig93reprint}: 
\begin{equation}\label{eq:strength_function}
\rho_W(E,k)=\sum_i |C_{ik}|^2\delta(E-E_i),
\end{equation}
which can be rewritten through the Green's function of the system  \cite{FG2000}:
\begin{equation}
G_{kj}(E)=\sum_i \frac{C_{ik}C^*_{ij}}{E-E_i+i\alpha},\quad \alpha>0,\  \alpha\rightarrow 0,
\end{equation}
\begin{equation}
\rho_W(E,k) = -\frac{1}{\pi}\mathrm{Im}[G_{kk}(E)].
\end{equation}
After performing an appropriate averaging to eliminate $C_{ik}$ fluctuations which take place in an individual $|\Psi_i\rangle$ the strength function can be expressed through certain self-energy operator $\Sigma_k$:
\begin{equation}\label{eq:general_lorenzian}
\overline{\rho_W(E,k)}=\frac{1}{2\pi}\frac{\Gamma_k}{(E-\varepsilon_k-\Delta_k)^2+\Gamma_k^2/4},
\end{equation}
\begin{equation}\label{eq:gen_lor_parameters}
\Gamma_k=-2\mathrm{Im}[\Sigma_k(E)],\quad \Delta_k=\mathrm{Re}[\Sigma_k(E)].
\end{equation}
Here $\Gamma_k$ is the energy spreading width of the basis component $k$ and $\Delta_k$ is the shift of the eigenvector center from the basis state energy $\varepsilon_k$. Generally speaking, $\Gamma_k$ and $\Delta_k$ depend on energy and the shape of \eqref{eq:general_lorenzian} doesn't have to be simply Lorenzian. In fact, in pure GOE it is a semicircle \cite{Wig93reprint,FG2000,GMW98}. But when the average squared off-diagonal element $\overline{V_{ij}^2}=V^2$ is not very large, allowing for the condition $V^2\ll D^2b$ ($D$ being average energy spacing between the basis states and $b$ the bandwidth of bRMT), the strength function for an infinite band random matrix can be written \cite{Wig93reprint,FG2000} as:
\begin{equation}\label{eq:lorenz_sf}
\overline{\rho_W(E,k)}=\frac{1}{2\pi}\frac{\Gamma}{(\varepsilon_k-E)^2+\Gamma^2/4}\ ,
\end{equation}
where the spreading width is now
\begin{equation}\label{eq:condit_lor}
\Gamma=\frac{2\pi V^2}{D}\ ,\quad\Gamma\ll Db\ .
\end{equation}
This approximation to the strength function is still applicable when both $V^2$ and $D$ change along the matrix, if the change is sufficiently slow. Nevertheless, in a real system the shift $\Delta_k\equiv-\Delta$ presented in \eqref{eq:general_lorenzian} should not be neglected in low-lying eigenstates. The repulsion of levels near the beginning of the energy spectrum is not compensated from below, therefore the resulting eigenvalue $E_i$ lies lower than the energies $\varepsilon_k$ of its basis states. For similar reasons  the shape of \eqref{eq:lorenz_sf} is asymmetrically distorted for the lower levels. In higher parts of the spectrum both of these edge effects decrease \cite{FGGK94}.

\begin{figure}[tb]
    \includegraphics[width=\columnwidth]{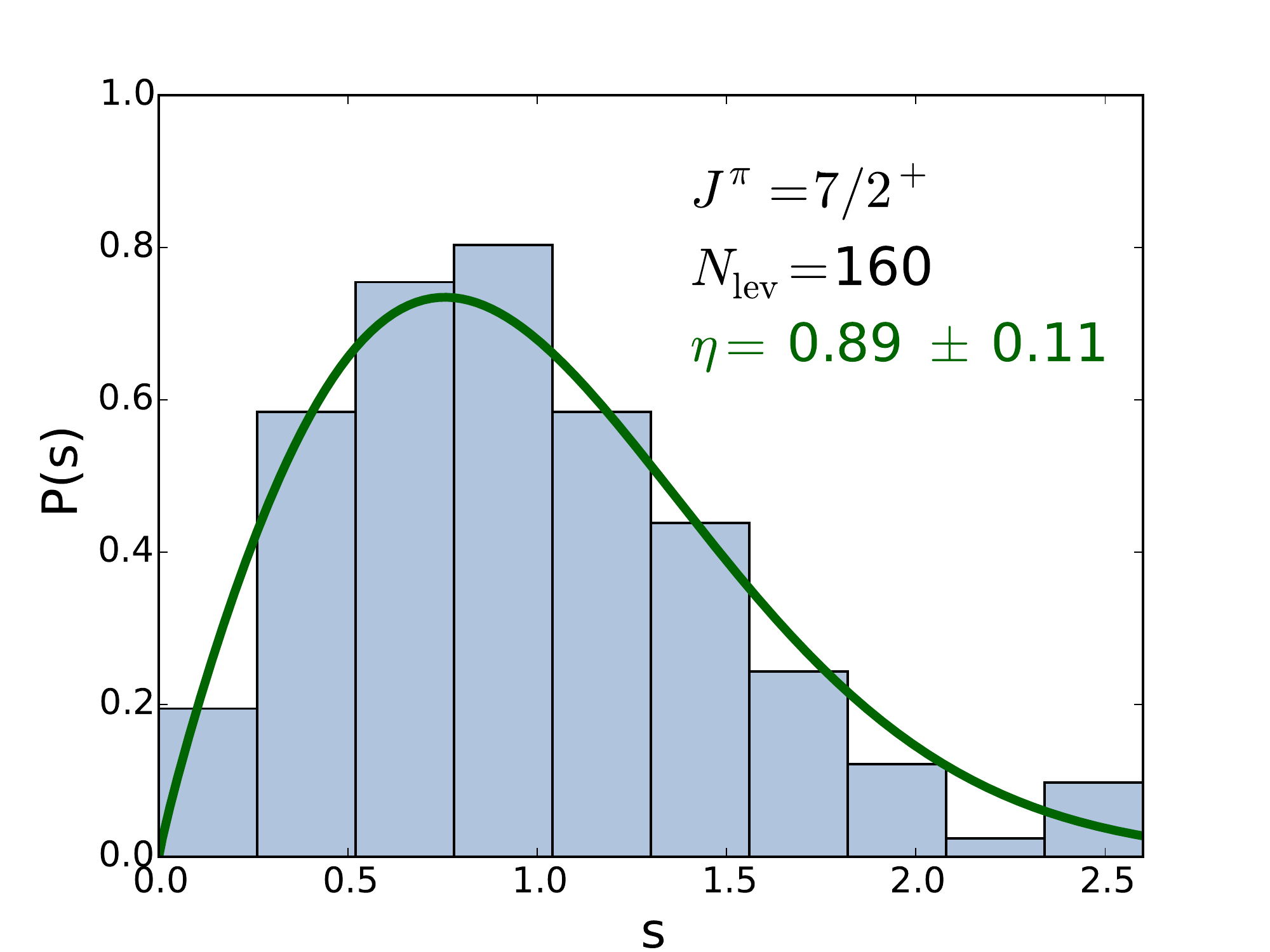}
    \caption{Nearest neighbor spacing (NNS) histogram built for the unfolded $J^\pi=7/2^+$ spectrum. Unfolded spacings probability distribution is fitted with Brody function (solid line, \eqref{eq:brody}) and repulsion parameter $\eta$ is obtained. $N_\mathrm{lev}$ is number of levels with $J^\pi=7/2^+$ considered during the fitting. Dimensionless level spacing $s$ is given in terms of local average spacing $D$ of the spectrum (see \sref{nns_distr}). Histogram is normalized to unity.}\label{fig:sample_hist}
\end{figure}

\begin{figure}[bt]
    \includegraphics[width=\columnwidth]{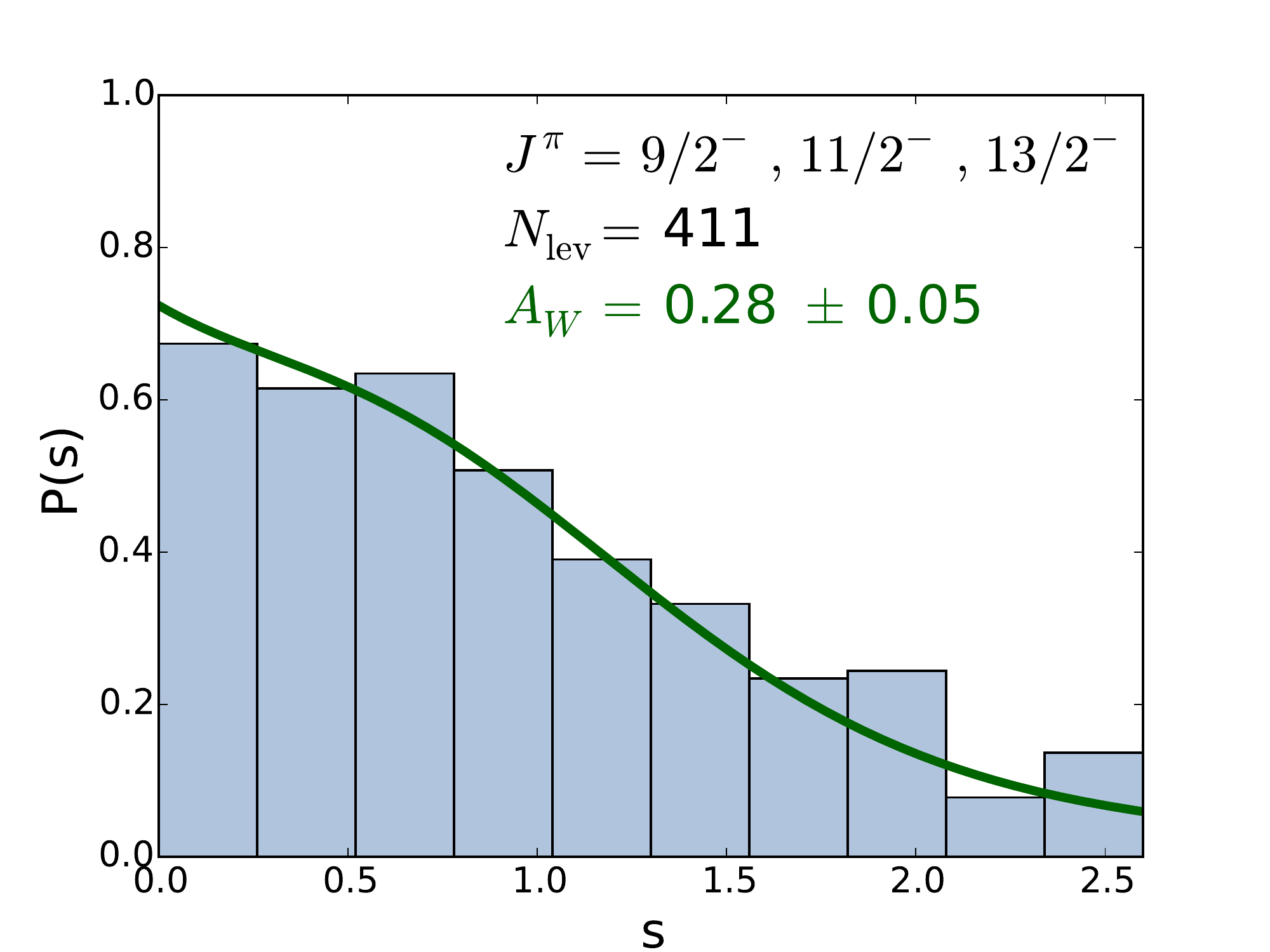}
    \caption{NNS statistics of $N_\mathrm{lev}=411$ levels accesible from the ground state $J^\pi=11/2^+$ with E1 transitions. The probability distribution of the spacings is fitted with a weighted sum of Wigner and Poisson distributions $p(S)=A_W p_\mathrm{Wigner}(s)+(1-A_W) p_\mathrm{Poisson}(s)$.
    }\label{fig:e1}
\end{figure}

\section{Results and discussion}
\subsection{NNS distribution histograms and fitting}
Using methods described in \sref{ci_model} and \sref{nns_distr} we obtained spacings distributions for different manifolds $J^\pi$. In each manifold only levels with $n>10$ are considered, because the first low-lying levels of the spectrum are not expected to participate in chaotic behavior. We examine energies $0\leq E_\mathrm{e}<44454$~cm$^{-1}$ for even eigenvalues and $8065$~cm$^{-1}<E_\mathrm{o}<40748$~cm$^{-1}$ for odd ones (zero corresponds to the ground state). The density of calculated levels on these energy ranges might be less than in a real Pa atom, since we do not account for core polarization in our model.

Probability density of spacings for each $J^\pi$ is estimated with Brody function \eqref{eq:brody}.  The resulting repulsion parameters $\eta$ for several manifolds $J^\pi$ are listed in \tref{tab:etas}, an example histogram of $J^\pi = 7/2^+$ is presented in \fref{fig:sample_hist}.

The results for repulsion parameters in \tref{tab:etas} are comparable with $\eta=1$ and thus with Wigner distribution for GOE. The only exception is $J^\pi=13/2^+$ where $\eta=0.64\pm 0.13$ suggests intermediate statistics between Poisson and Wigner cases. It can be due to slightly lower density of this spectral manifold: there are only $N+10=93$ levels of $J^\pi=13/2^+$ on the energy range $0<E_\mathrm{e}<44454$~cm$^{-1}$.

\begin{table}[bth]
    \centering
    \begin{tabular}{|c|r|r||r|r|}
        \hline
        & \multicolumn{2}{c||}{$\pi=+1$} & \multicolumn{2}{c|}{$\pi=-1$}\\
        \hline
        $J$ & $N_\mathrm{lev}$ & $\eta$ & $N_\mathrm{lev}$ & $\eta$ \\
        \hline
        $7/2$  & 160 & $0.89\pm 0.11$ & 151 & $0.94\pm 0.12$\\
        $9/2$  & 153 & $0.96\pm 0.12$ & 156 & $0.91\pm 0.11$\\
        $11/2$ & 121 & $0.92\pm 0.12$ & 136 & $0.99\pm 0.13$\\
        $13/2$ & 83 &  $0.64\pm 0.13$ & 99 &  $0.95\pm 0.15$\\
        \hline
    \end{tabular}
    \caption{Repulsion parameters $\eta$ obtained from fitting NNS data with Brody function \eqref{eq:brody}. Each set of spacings corresponds to a spectral subspace with fixed parity $\pi$ and total angular momentum $J$. The number of levels of a given $J^\pi$ manifold participating in the fitting is denoted as $N_\mathrm{lev}$.}\label{tab:etas}
\end{table}

\begin{figure}[htb]
    \includegraphics[width=\columnwidth]{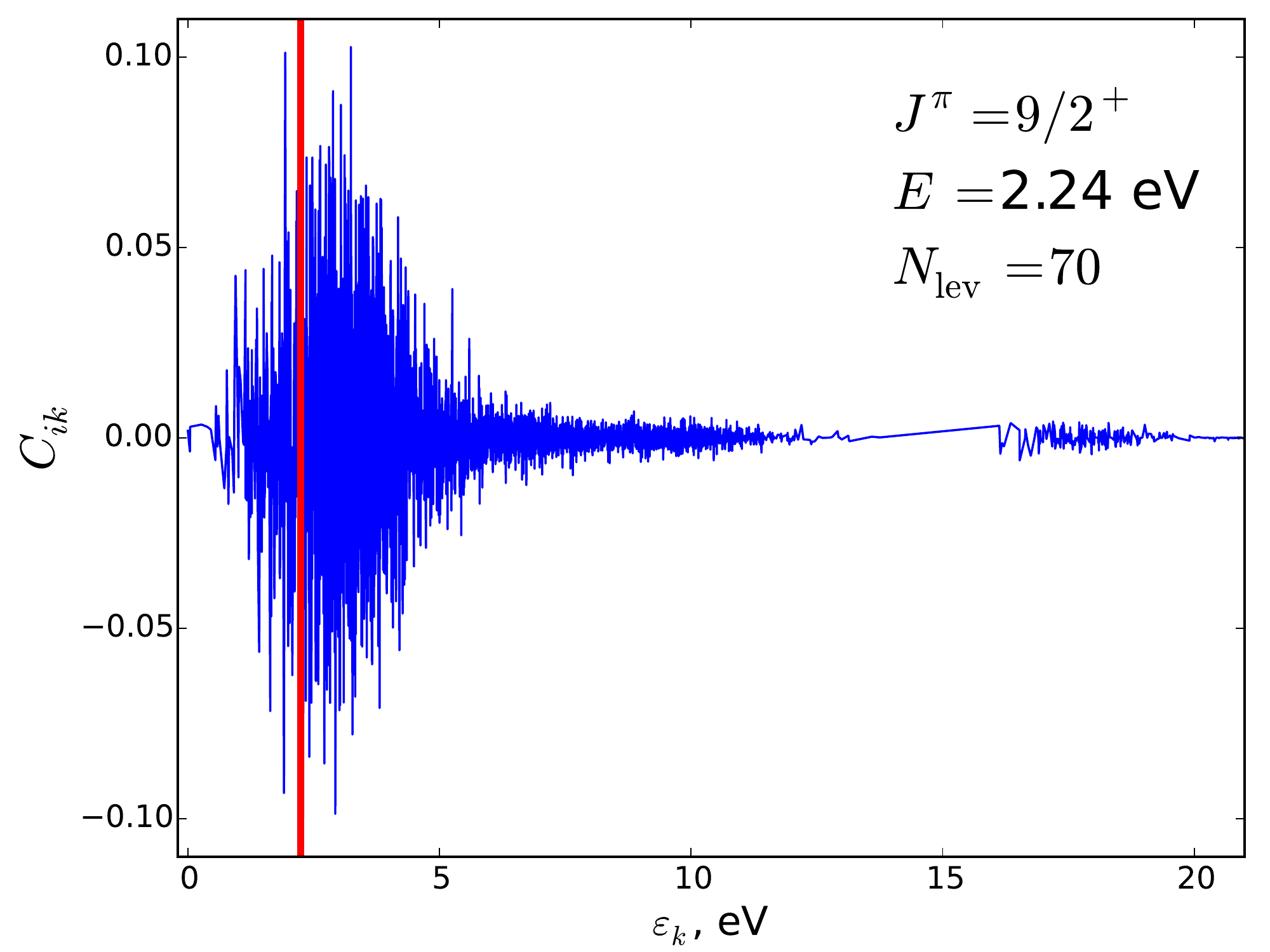}
    \caption{Coefficients $C_{ik}=C_i(\varepsilon_k)$ arranged by the basis state energy $\varepsilon_k$. The plot corresponds to the eigenfunction with energy $E=0.08247$~a.u.~$=2.24418$~eV (vertical line), which has the succesive number $N_\mathrm{lev}=70$ in manifold $J^{\pi}=9/2^+$. We define the lowest basis state energy $\varepsilon_{k0}$ as being zero. Large basis components lie within a certain energy interval. Admixture of isolated components can be considered small. Therefore we can define an energy spreading width $\Gamma$ for the given eigenfunction. The coefficients $C_{ik}$ behave like random variables with the variance $\langle C_{ik}^2\rangle$ depending on the energy difference $(\varepsilon_k- E_i)$ - see \fref{fig:c2} and Eqns.\eqref{eq:fitlor}, \eqref{eq:sqlor}.}\label{fig:c}
\end{figure}

\begin{figure}[htb]
    \includegraphics[width=\columnwidth]{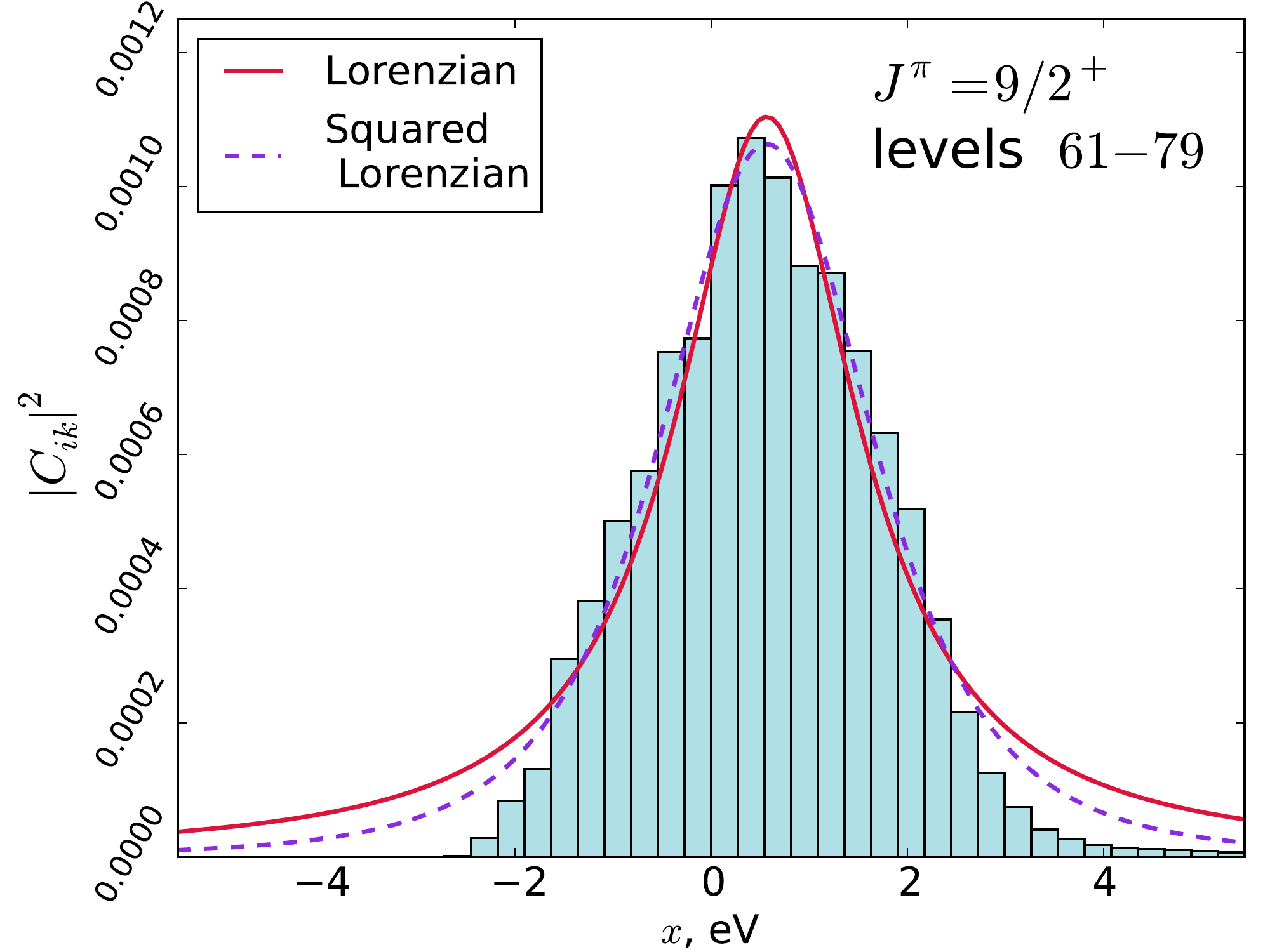}
    \caption{
        Averaged binned statistics of squared coefficients $|C_{ik}|^2=|C_i(\varepsilon_k)|^2$ for 61--79 levels of $J^{\pi}=9/2^+$, where we assume $x=\varepsilon_k-E_i$ being the difference between basis state energy $\varepsilon_k$ and the eigenvalue $E_i$. Lorenzian \eqref{eq:fitlor} and squared Lorenzian \eqref{eq:sqlor} fits are applied, with estimated overall shift $\Delta\approx 0.57$~eV. The resulting parameters are spreading width $\Gamma=2.2\pm 0.1$~eV and number of principal components $N=905\pm 38$ for the Lorenzian, $\Gamma=2.0\pm 0.1$~eV and $N=939\pm 23$ for the squared Lorenzian fit.
    }\label{fig:c2}
\end{figure}

\subsection{Spreading width}

The strength function \eqref{eq:strength_function} is connected to the smooth envelope $w(\varepsilon_k,E)$ of the squared coefficients $|C_{ik}|^2=|C_{i}(\varepsilon_k)|^2$ as follows \cite{FGGK94}:
\begin{equation}
\overline{\rho_W(E,k)}=D^{-1}\overline{|C_{ik}|^2}\equiv D^{-1}w(\varepsilon_k,E)\ ,
\end{equation}
with the local mean level spacing defined as:
\begin{equation}
D^{-1}\equiv\overline{\rho_\mathrm(E)}=\overline{\sum_i \delta(E-E_i)}\ ,\quad E\simeq E_i\ .
\end{equation}
In our case, averaging is performed on the neighboring eigenstates to account for possible gradual change of the smooth envelope along the spectrum.
It is supposed that the Hamiltonian matrix of Pa fulfills the conditions for its strength function to be roughly of Lorenzian shape \eqref{eq:lorenz_sf}, with the addition of a possible overall shift $\Delta$. Therefore, we consider levels far enough from the edges of the spectrum. Number of principal components is introduced as $N\equiv\frac{\pi\Gamma}{2D}$. Then the Lorenzian smooth envelope of the squared coefficients could be written:
\begin{align}
w(\varepsilon_k,E)&=D\overline{\rho_W(E,k)}\\ &=\frac{1}{N}\frac{\Gamma^2/4}{(\varepsilon_k-E-\Delta)^2+\Gamma^2/4}\label{eq:fitlor}\ .
\end{align}
For averaging over neighboring levels, it is convenient to treat $x=\varepsilon_k-E$ as a single variable; we presume that the shift $\Delta$ is constant for close eigenvalues. First, we make binned statistics for $x=[-0.2,0.2]$~a.u. $\approx[-5.44,5.44]$~eV and 40 bins. Then each bin is averaged over 19 neighboring levels. The resulting binned plot is fitted by the Lorenzian \eqref{eq:fitlor}. \tref{tab:c2} contains estimated parameters and \fref{fig:c2} is an example of fitted histogram-like plot.

It should be noted that the tails of the plot decrease much faster than predicted by \eqref{eq:fitlor}, since the condition $|\varepsilon_k-E|<Db$ corresponding to \eqref{eq:condit_lor} is being violated and the tails start to drop exponentially \cite{Wig93reprint,FGGK94, CCGI96}. For comparison, we use another function for fitting that decreases faster on the edges than \eqref{eq:fitlor}, namely the squared Lorenzian:
\begin{equation}
w'(\varepsilon_k,E)=\frac{1}{N}\frac{(\tilde{\Gamma}^2/4)^2}{[(\varepsilon_k-E-\Delta)^2+\tilde{\Gamma}^2/4]^2}\ ,\ \Gamma=\frac{\tilde{\Gamma}}{2}\label{eq:sqlor}.
\end{equation}
Spreading widths $\Gamma$ and numbers of principal components $N$ obtained from fitting $w'(\varepsilon_k,E)$ agree with those resulting from the Lorenzian fit \eqref{eq:fitlor}. The shift $\Delta$ was estimated as the expected (mean) value of the binned plot.

\begin{table}[bt]
    \centering
    \begin{tabular}{|r|r|r|r|r|}
        \hline
        $J$ & levels & $\Gamma$, eV & $N$ & $\Delta$, eV \\
        \hline
        \multicolumn{5}{|c|}{$\pi=+1$}\\
        \hline
        $7/2$   &   61--79  & $ 2.4 \pm 0.1 $ & $   871 \pm 36  $ & $   0.35\pm 0.14    $ \\
        $7/2$   &   111--129    & $ 2.7 \pm 0.2 $ & $   1171    \pm 57  $ & $   0.12\pm 0.14    $ \\
        \hline
        $9/2$   &   61--79  & $ 2.2 \pm 0.1 $ & $   905 \pm 38  $ & $   0.57\pm 0.14    $ \\
        $9/2$   &   111--129    & $ 2.6 \pm 0.2 $ & $   1192    \pm 62  $ & $   0.37\pm 0.14    $ \\
        \hline
        $11/2$  &   61--79  & $ 2.2 \pm 0.2 $ & $   996 \pm 47  $ & $   0.71\pm 0.14    $ \\
        \hline
        $13/2$  &   60--78  & $ 2.2 \pm 0.1 $ & $   1106    \pm 47  $ & $   0.65\pm 0.14    $ \\

        \hline
        \multicolumn{5}{|c|}{$\pi=-1$}\\
        \hline
        $7/2$   &   61--79  & $ 1.7 \pm 0.1 $ & $   955 \pm 30  $ & $   0.64\pm 0.14    $ \\
        $7/2$   &   111--129    & $ 2.4 \pm 0.2 $ & $   1706    \pm 88  $ & $   0.35\pm 0.14    $ \\
        \hline
        $9/2$   &   61--79  & $ 1.9 \pm 0.1 $ & $   1124    \pm 46  $ & $   0.78\pm 0.14    $ \\
        $9/2$   &   111--129    & $ 2.4 \pm 0.2 $ & $   1825    \pm 91  $ & $   0.58\pm 0.14    $ \\
        \hline
        $11/2$  &   61--79  & $ 2.0 \pm 0.1 $ & $   1483    \pm 58  $ & $   0.96\pm 0.14    $ \\
        \hline
        $13/2$  &   61--79  & $ 2.3 \pm 0.1 $ & $   2124    \pm 87  $ & $   1.12\pm 0.14    $ \\
        \hline
    \end{tabular}
    \caption{Least-squares parameters of the Lorenzian fit \eqref{eq:fitlor}. $J$ and $\pi$ are the total angular momentum and parity of the given manifold of wave functions, the second column shows the numbers of $J^\pi$ levels participating in averaging of coefficients $|C_i(\varepsilon_k)|^2$. Spreading width $\Gamma$ and number of principal components $N$ are listed in the next two columns. Approximate shift $\Delta$ of the Lorenzian fit with respect to the eigenvalue is considered constant over averaged levels; its error is estimated as half size of the bin. The shift decreases in the higher part of the spectrum.}\label{tab:c2}
\end{table}

\subsection{Small perturbation enhancement}\label{sec:small_pert}

One of the most important features of the chaotic systems, as we
mentioned in the Introduction, is high sensitivity to small
perturbations. In particular, the mixing of the states scales with
the number of principal components $N$ in the wave function as
\eqref{sqrtN}. This scaling holds only while the mixing is small,
but for sufficiently large $N$ the mixing becomes strong.
Perturbation at this point can not be considered small anymore and
perturbation theory fails.

\begin{figure}[htb]
    \includegraphics[width=\columnwidth]{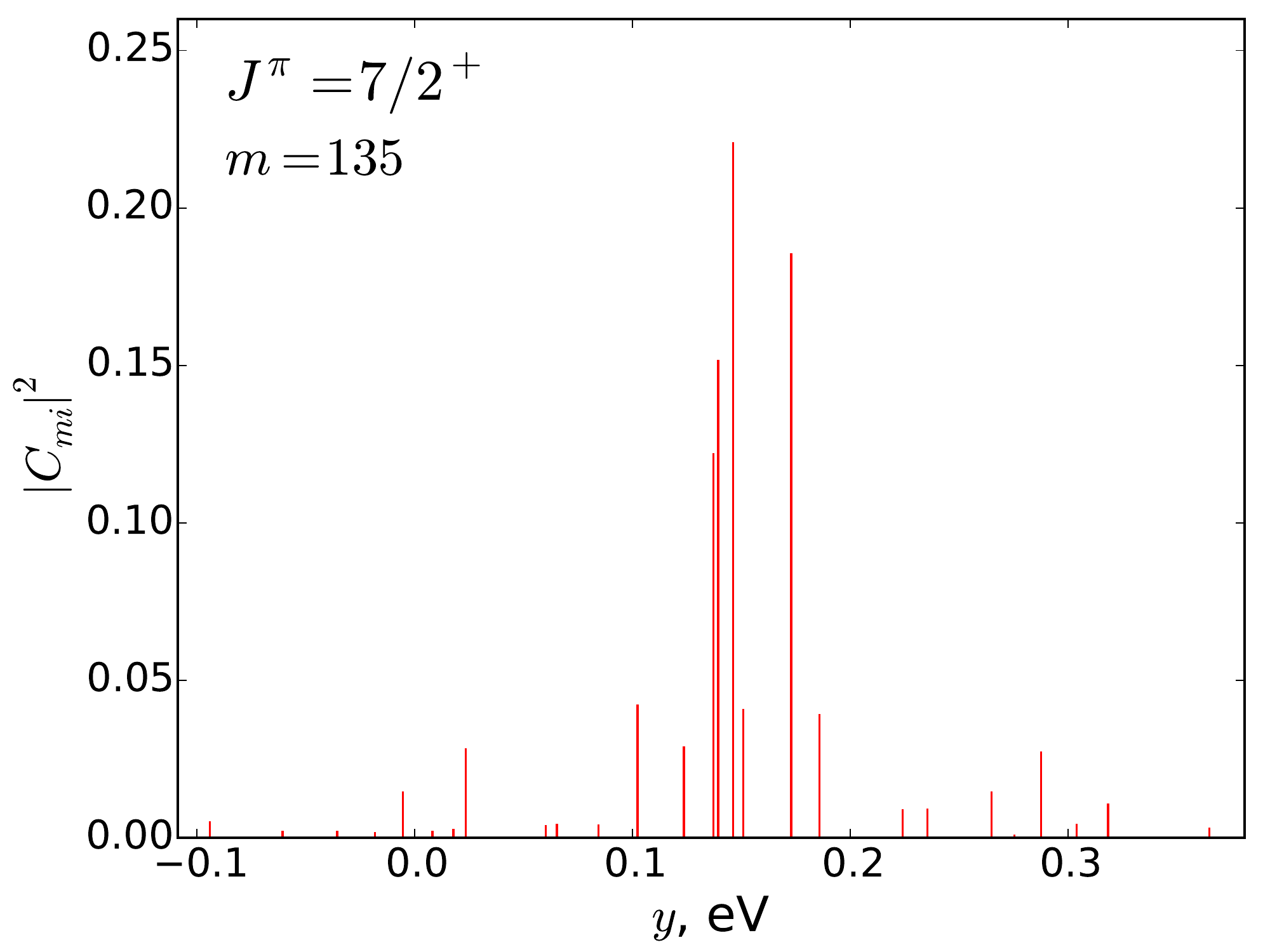}
    \caption{Weights of the eigenfunctions of the two-particle Hamiltonian
    in the eigenfunction of the Hamiltonian with included TEI. The plot is
    for eigenfunction number 135 from the subspace $7/2^+$. We denote $y=E_{0,i}-E_k$. The energies
    $E_{0,i}$ and $E_k$ correspond to the unperturbed and full Hamiltonians respectively. }
    \label{fig_3e_wf_45}
\end{figure}

As an example of the behavior described above, we have studied the
effective three-electron interaction (TEI) between valence
electrons. Such interaction is caused by the core polarization
effects \cite{DFSS91,DFK96}. Typically it is very small, about $10^{-3}$ of
the residual Coulomb interaction between valence electrons. The
latter is defined as the difference between the two-electron Coulomb
interaction and the self-consistent field, used to form the
one-electron orbitals. Residual interaction determines
configurational mixing. For atoms and ions with filling $d$, or  $f$ shells
TEI is enhanced by one or two orders of magnitude \cite{BFK08,KSPT16}, but
is still much smaller than the residual interaction, which is
typically of the order of unity in atomic units.

We calculated TEI in protactinium for the subspace $7/2^+$
with one of the highest level densities. The average ratio of the
TEI and the residual Coulomb non-diagonal matrix elements is found
to be 0.017. Similarly, the ratio for the maximal matrix elements is
equal to 0.015. Thus, for the non-chaotic system one would expect
rather small mixing of the eigenfunctions. However, diagonalization
of the TEI Hamiltonian for the subspace $7/2^+$ results in a
complete mixing of the unperturbed eigenfunctions. An example of one
of the new eigenfunctions in the basis set of the old ones is shown
in \fref{fig_3e_wf_45}. We see that there are 4 principal components
with comparable weights and about 15 components with weights above
1\%. This means that effective three-electron interaction in
protactinium can not be considered small and has to be treated on
the same footing as the residual Coulomb interaction. This result is
in agreement with the estimate \eqref{sqrtN}. For the subspace
$7/2^+$ the number of principal components is $N\sim 10^3$.
Multiplying the ratio of the matrix elements by $\sqrt{N}\sim 30$ we
get $0.017\cdot 30=0.5$.

In fact, estimate \eqref{sqrtN} gives only the lower limit of possible scaling, since it is written for a single-particle interaction $V$. Systems with multiple-particle interaction $V_\mathrm{mult}$ can bear additional factor $M$ equal to the number of non-zero matrix elements between the basis states $\langle\Phi_i|V_\mathrm{mult}|\Phi_j\rangle$ with a fixed $i$.
\begin{equation}
\frac{\langle\Psi_a|V_\mathrm{mult}|\Psi_b\rangle}{\Delta E_{ab}}\sim \sqrt{M(N)\cdot N},
\end{equation}
\begin{equation}
1<M(N)\ll N\ .
\end{equation}
Therefore multiple-particle interaction mixings can scale faster than $\sqrt{N}$. In the case of $n=5$ valence electrons, a three-particle operator and $N\approx 1000$, the factor can be estimated as $M(N)\approx 15$.

\section{Conclusion}

According to the CI calculation described in \sref{ci_model} Pa atom clearly shows many-body
chaos behavior in its energy spectrum, starting already from relatively
close to the ground level. Properties of two-electron Hamiltonian matrix of Pa correspond to those of random
two-body interaction (RTBI) matrices which have large leading
diagonal and sparse band-like structure of random interaction
non-diagonal elements. RTBI model demonstrates some aspects of behavior close to pure random matrix theory, such as Wigner distribution of
spacings between energy levels, but it differs, for instance, in the
composition of its eigenfunctions \cite{FIC96, FW71, BF71a}. In this
regard Pa atom is very similar to highly chaotic Ce atom thoroughly
investigated before \cite{FGGK94,FGGP99,GFG95}; therefore properties of its Hamiltonian can be treated statistically \cite{FGGP98}. Approximate quantum numbers such as total electron orbital angular momentum $L$ and spin $S$ disappear due to the enhancement of the spin-orbit interaction and such classification of atomic energy levels, which is still present in
the Tables, becomes meaningless \cite{FGGK94,FGGP98,FGGP99,GFG95}. Number of principal components
participating in excited eigenstates of Pa is $N\sim 10^3$, an order
of magnitude larger than for Ce and closer to that of compound
nuclei ($N\sim 10^4 - 10^6$). 

Such strong mixing of basis states is of particular interest, since it leads to statistical enhancement of small perturbations, another signature of quantum many-body chaos. This fact was illustrated by the calculation of effective three-electron interaction of unperturbed Hamiltonian eigenstates in \sref{sec:small_pert}. The mixing turned out to be close in strength to the residual Coulomb interaction mixing already accounted for in the unperturbed Hamiltonian, confirming drastic enhancement of a small interaction. 

Other small perturbations can be enhanced in a similar fashion and made feasible for experimental observation, e.g. parity non-conserving mixings due to weak interaction between the atomic nucleus and electrons.  

In conclusion, we would like to note that an indication of chaos in the spectra of Pa atom near the ionization threshold has been recently observed by the Prof. Wendt group at Johannes Gutenberg University, Mainz \cite{NW16_unpubl}.

\section*{Acknowledgments}

This work is partly supported by Russian Foundation for Basic
Research Grant No.\ 14-02-00241, the Dynasty Foundation Scholarship, the Australian Research Council and the Gutenberg Fellowship of Johannes Gutenberg University, Mainz. Authors want to thank Prof. Klaus Wendt and Pascal Naubereit for stimulating discussion. M.G.K. would like to thank Mainz Institute for Theoretical Physics and Helmholtz Institute Mainz for hospitality.

\bibliographystyle{apsrev} 

\end{document}